\documentclass[aps,prb,twocolumn,showpacs]{revtex4}
\usepackage{graphicx}
\begin{document}

\title{Two dimensional electron gas driven by strong alternating electric field}

\author{A. Kashuba}

\affiliation{L. D. Landau Institute for Theoretical Physics, Russian Academy of Sciences, 2 Kosygina str., 119334 Moscow}

\date{\today}

\begin{abstract}
2D Fermi liquid driven by uniform \textit{ac}-electric field at zero temperature may remain in quantum coherent non-equilibrium state. We develop a quasistatic approximation for strong and slow \textit{ac}-fields and solve the problem of driven disordered 2D electron gas in high non-overlapping Landau levels. The broadening of Landau level has the Lorentz form and is inversely proportional to the amplitude of \textit{ac}-field. In the absence of electron-phonon interaction the electron distribution function is constant within the last Landau level and the diagonal \textit{dc}-conductivity is zero. For weak electron-phonon interaction the \textit{dc}-conductivity is anisotropic. A kinetic transition from the phonon cooling to the phonon heating is predicted.
\end{abstract}


\maketitle

\section{Introduction}

Experience with microwave ovens teaches us that electron systems subjected to an alternating (\textit{ac}) electric field are heating up. This paper explores examples to the contrary. A closed system remains in a quantum coherent state even if driven by \textit{ac}-field. This driven state is equivalent to the static ground state provided the system possesses the translational invariance, the parabolic band dispersion and instantaneous electron-electron interaction. Boundaries and disorder preserve the coherence but profoundly transform the driven state. It is the electron-phonon interaction that destroys the coherence and eventually heats up the electron system. At low temperature the electron-phonon interaction is especially weak and the quantum coherence of a driven system can be observed. In three dimensions the \textit{ac}-electric field is confined near the surface but in two dimensions (2D) it penetrates and drives electron system into a new state. 

Recently in high mobility GaAs heterostructures under microwave irradiation in perpendicular magnetic field new oscillations of conductivity and new zero-resistance states has been discovered \cite{Mani,zdpw}. These oscillations depend on the ratio of the \textit{ac}-frequency $\omega$ to the cyclotron frequency $\omega_H$. Ryzhii has predicted a phenomenon of the absolute negative conductivity in two dimensional electron gas (2DEG) driven by the microwave field \cite{ryzh} and, recently, his theory has been confirmed in several models \cite{dsrg,ll03} of disordered electron system.

Proper description of the absolute negative conductivity requires to account for a current necessitated by electron relaxation from the exited Landau levels predominantly by emitting phonons \cite{ryzh03}. This compensating current amounts to a positive conductivity. In GaAs the phonon effect is weaker than the Ryzhii effect of disorder and the theory neglects phonons in the limit $\omega_H\ll E_{ph}$, where $E_{ph}$ is the characteristic phonon energy (5 Kelvins in GaAs).

The problem of disordered 2DEG in magnetic fields that quantize the electron motion is a special one because it is the disorder that lifts the Landau level degeneracy and imparts a velocity to electrons, and it is the disorder that is coupled to the \textit{ac}-field. The quantum limit corresponds to a weak disorder: $\omega_H\tau\gg 1$, where $\tau$ is the elastic scattering time. In high Landau levels the exact self-consistent Born approximation theory predicts the semi-circle density of states and longitudinal resistivity proportional to the magnetic field \cite{AFS}. 

In this paper I find modifications in this picture due to microwaves in the limit of slow and strong \textit{ac}-field $E$: $\omega\ll \omega_H$ and $\omega_H^3\ll (eEv_F)^2\tau$, where $v_F$ is the Fermi velocity, called a quasistatic approximation. In strong fields an electron drifts with large velocity 'over the disorder' and experience predominantly forward scattering \cite{PG}. Accordingly the longitudinal \textit{dc}-conductivity of the driven 2DEG is strongly suppressed and is zero in the absence of the electron-phonon interaction. The semi-circle density of states of the static 2DEG transforms into the Lorentz spectral density of the driven 2DEG with the broadening of Landau levels being inversely proportional to $E$. 

In the first two sections we overview a Fermi liquid driven by \textit{ac}-field. Here we set a stage by examining examples from the usual Fermi liquid. In particular a quasistatic approximation and a special driven state of 1D Fermi gas with boundaries that possesses a smooth electron distribution function are introduced. In Sec. III we find a kinetic transition from the phonon cooling to the phonon heating. In the last two sections we solve the problem of the \textit{ac}-driven disordered 2DEG in high non-overlapping Landau levels with Sec. V containing the main results.

\section{Fermi liquid in \textit{ac}-electric field}

In this section we overview the theory of Fermi liquid driven by uniform \textit{ac}-electric field. We determine the quasiparticle dispersion, the linear response function and effects due to interactions with phonons and impurities. 

Uniform in-plane electric field $\vec{E}(t)$ can be described in the gauge with zero electrostatic and uniform vector potential $\vec{A}(t)=-c\int \vec{E}(t)dt$ assumed to be limited and, therefore, with no arbitrary gauge degrees of freedom. We write the Hamiltonian of 2D Fermi liquid $\hat{H}(t)=\hat{H}_0[\vec{A}(t)]+\hat{H}^{int}$ as:
\begin{equation}\label{Ham}
\hat{H}(t)=\frac{1}{2m}\sum_i\left(-i\vec{\nabla}_i+\frac{e}{c}\vec{A}(t) \right)^2 + \sum_{i<j} U(\vec{r}_i-\vec{r}_j),
\end{equation}
and put $\hbar=1$ throughout this paper. Classically a free electron oscillates in \textit{ac}-field with the velocity: $\vec{V}(t)=-e\vec{A}(t)/mc$. The wave-function $\Psi(t)$ of a closed system is a solution of the Schroedinger equation:
\begin{equation} \label{SchrEq}
i\frac{\partial}{\partial t}\Psi(t)= \hat{H}(t)\Psi(t).
\end{equation}
If the Hamiltonian does not depend on time then its eigenvalue - the energy - is conserved. The momentum operator commutes with the one-particle Hamiltonian $\hat{H}_0[\vec{A}]$ and the momentum of electron $\vec{p}$ is a good quantum number. The solution of one-particle Schroedinger equation with momentum $\vec{p}$ yields:
\begin{equation} \label{FloqWF}
\psi_{\vec{p}}(t,\vec{r}) =\frac{e^{i\vec{p}\vec{r}}}{\sqrt{S}} \exp\left(\frac{-i}{2m} \int_{-\infty}^t \left(\vec{p}+\frac{e}{c}\vec{A}(t)\right)^2\ dt\right),
\end{equation}
where $S$ is the area of sample. If $\vec{A}(t)$ is a periodic function of time with the period $2\pi/\omega$ then apart from the multiplicative factor $\exp(-i\tilde{\epsilon}t)$ the electron states (\ref{FloqWF}) - known as Floquet eigenstates - are periodic in time whereas $\tilde{\epsilon}=p^2/2m\ {\rm mod}\ \omega$ ($0\le \tilde{\epsilon}<\omega$) is a quasienergy.

We expand the Hamiltonian (\ref{Ham}) in powers of $\vec{A}(t)$:
\begin{equation}
\hat{H}(t)=\hat{H}_0+\hat{H}^{int}-\frac{1}{c}\vec{A}(t) \hat{\vec{J}}_0+ \frac{e^2}{2mc^2}\vec{A}^2(t) \hat{N}_0,
\end{equation}
where $\hat{H}_0=\hat{H}_0[0]$ and $\hat{H}_0+\hat{H}^{int}$ is the static Hamiltonian, $\vec{J}_0= (ie/mc)\sum_i \vec{\nabla}_i$ and $\hat{N}_0=\sum_i 1$ are the total electron current and number operators. All these three operators commutes with each other. Therefore the many-body wave function can be chosen to be the eigenfunction of all these three operators with the total energy $E$, the total number of electrons $N$ and the total momentum $\vec{P}$ and can be expanded in the basis of the momentum eigenfunctions: $\psi_{\vec{p}}(\vec{r})= \exp(i\vec{p}\vec{r})/\sqrt{S}$,:
\begin{equation} \label{MbWF}
\Psi_0=e^{-iEt}\!\! \sum_{\vec{p}_1..\vec{p}_N} a(\vec{p}_1..\vec{p}_N) \prod_{i=1}^N \psi_{\vec{p}_i}(\vec{r}_i),
\end{equation}
where $a(\vec{p}_1..\vec{p}_N)$ are complex-valued coefficients that are non-zero if $\sum_i\vec{p}_i=\vec{P}$. The wave function:
\begin{equation} \label{MbWFn}
\Psi(t)=e^{-iEt}\!\! \sum_{\vec{p}_1..\vec{p}_N} a(\vec{p}_1..\vec{p}_N) \prod_{i=1}^N \psi_{\vec{p}_i}(t,\vec{r}_i) e^{itp_i^2/2m},
\end{equation}
where $\psi_{\vec{p}_i}(t,\vec{r}_i)$ is given by Eq.(\ref{FloqWF}), is the solution of the time dependent Schroedinger equation (\ref{SchrEq}) with the Hamiltonian (\ref{Ham}). It is the eigenfunction of the total electron current and number operators: $\hat{J}_0\Psi(t)= e\vec{P}/mc\ \Psi(t)$ and $\hat{N}_0\Psi(t)=N\Psi(t)$. The solution (\ref{MbWFn}) requires the same conditions as the Kohn theorem \cite{Kohn}: the electron band dispersion is strictly quadratic, the electron-electron interaction is instantaneous and the system is translationally invariant without boundaries or bulk disorder. Weak interactions violating these conditions - boundary, disorder or the electron-phonon interaction - can be treated in perturbation theory \cite{ll03}. 

Canonical transformation: $\hat{U}^\dagger(t)[i\partial/\partial t- \hat{H}(t)] \hat{U}(t)= i\partial/\partial t- \hat{H}_0-\hat{H}^{int}$, of the second quantized electron operators: $\psi(\vec{p})=U(t,\vec{p}) \chi(\vec{p})$, where
\begin{equation}\label{Canonical}
U(t,\vec{p})=  \exp\left(-i\frac{e}{mc}\int_{-\infty}^t \left(\vec{p}\vec{A}(t)+\frac{e}{2c} \vec{A}^2(t)\right) dt\right),
\end{equation}
eliminates the \textit{ac}-electric field from the Schroedinger equation. Essentially it is a transformation from the laboratory frame to the center of mass frame oscillating with electrons \cite{ll03}. If a stimulus is canonically transformed then the response is the same as in the static system.

The electron density operator $\hat{\rho}(t,\vec{r})= \hat{\psi}^\dagger(\vec{r}) \hat{\psi}(\vec{r})$ in the driven 2DEG can be expressed in terms of the static density operator by using the wave function (\ref{FloqWF}):
\begin{equation} \label{Density}
\hat{\rho}(t,\vec{q})=\hat{\rho}_0(\vec{q}) \exp\left( i\vec{q}\int_{-\infty}^t \vec{V}(t) dt\right)
\end{equation}
where $\hat{\rho}_0(\vec{q})= \sum_{\vec{p}} \hat{\psi}^{\dagger}(\vec{p}) \hat{\psi}(\vec{p}+\vec{q})$ is the static density operator. Therefore, a generic instantaneous translationally invariant electron-electron interaction: $\hat{H}^{int}=\frac{1}{2} \sum_{q} \hat{\rho}_0(\vec{q}) U(\vec{q}) \hat{\rho}_0(-\vec{q})$, is invariant under the canonical transformation (\ref{Canonical}): $\hat{H}^{int}= \frac{1}{2} \sum_{q} \hat{\rho}(t,\vec{q}) U(\vec{q})\hat{\rho}(t,-\vec{q})$.

In the static Fermi liquid the quasiparticle dispersion reads: $\xi(\vec{p})= v_F(|\vec{p}|-p_F)$,\cite{AGD} where $p_F$ is the Fermi momentum related to electron density. In the driven Fermi liquid the Floquet theorem and the momentum conservation allows for two quasiparticles from the inside of the Fermi circle to scatter into empty states outside the Fermi circle with the total energy increment being an integer-multiple of $\omega$. But the existence of canonical transformation (\ref{Canonical}) shows that the matrix elements of such events are zero and that the quasienergy of quasiparticle is the same: $\tilde{\xi}(\vec{p}) = v_F(|\vec{p}|-p_F)$. In Landau theory the relaxation of quasiparticles with momenta close to $p_F$ is due to the electron-electron interaction and in both static and driven Fermi liquids $\tau_{in}\sim(p-p_F)^{-2}$.

\textbf{Incoherent driven state.} The static ground state is transformed into a driven state during adiabatic switching of the \textit{ac}-field at zero temperature. This driven state is no longer the unique lowest energy state because the energy is no longer conserved. In the absence of the electron-phonon interaction the driven state is a quantum coherent state but weak electron-phonon interaction directs the driven system into an incoherent state characterized by an effective electron temperature $T^*$. The electron-electron scattering probabilities are the same in both driven and static Fermi liquids. Therefore the solution of the Boltzmann kinetic equation with the electron-electron collision integral yields the Fermi-Dirac distribution function $f(\epsilon/T^*)$ that depends on the eigenvalue of the static Hamiltonian $\epsilon$ and some $T^*$ conserved by the electron-electron interaction, despite the bulk being kept at $T=0$. $T^*$ is determined from the balance of two energy flows: the 'heating' disorder scattering and the 'cooling' phonon emission.

\textbf{Coherent quantum state of driven system with boundary} emerges in the absence of electron-phonon interaction. Consider 1D non-interacting electron gas driven by parallel \textit{ac}-field and confined to a box of length $L$ with two boundaries of zero transparency. Electron states near the Fermi points $\pm p_F\gg 1/L$ are the sum of wave functions (\ref{FloqWF}) with left and right moving components:
\begin{equation}\label{Boundary}
\Psi=\sum_{p>0}\left( A(p) e^{i p(x-X(t))}+B(p) e^{-i p(x-X(t))} \right) e^{-i\frac{p^2}{2m}t},
\end{equation}
where $X(t)=\int V(t) dt$ and the \textit{ac}-field is periodic: $V(t)=V\sin(\omega t)$. Floquet theorem states that momenta in Eq.(\ref{Boundary}) are quantized: $p_n^2= p_0^2 +2mn\omega$, where $p_0$ is the middle momentum and $n$ is integer. Corresponding amplitudes $A_n=A(p_n)$ and $B_n=B(p_n)$ are related as $A_n=(-1)^{n+1}B_n$. Two boundary conditions: $\Psi|_{x=-L/2}(t)=0$ and $\Psi|_{x=L/2}(t)=0$, can be rewritten as: $\sum_k M_{nk}A_k=0$, with the matrix
\begin{equation}\label{BoundDis}
M_{nk}=i^{n-k}J_{n-k}\left(\frac{p_k}{\omega} V\right)\sin\left(\frac{p_k L}{2} +\frac{\pi n}{2}\right),
\end{equation} 
where $J_n(x)$ is the Bessel function. $p_0$ is found from the self-consistency equation $\det{M}(p_0)=0$. We solve numerically for $A_n$ and find a broad electron distribution over the energy range: $\delta \epsilon\sim n_{max}\omega\sim V p_F$. A Slater determinant constructed from all wave functions with $|p_0|<p_F$ is the quantum coherent state that turns into the static Fermi gas at $V=0$ and that describes a smooth electron distribution in the momentum space. 

\textbf{Linear response function} of charged Fermi gas is the response to the scalar potential $\phi(t,\vec{r})$. If the \textit{ac}-field velocity is periodic: $\vec{V}(t)=\vec{V}\sin(\omega t)$, then the response function: 
\begin{equation} \label{resp}
\langle \hat{\rho}(t,\vec{q})\rangle=\chi(t,t',\vec{q})\phi(t',\vec{q}),
\end{equation}
is also periodic: $\langle \hat{\rho} (\Omega+n\omega,\vec{q})\rangle =\chi_n(\Omega,\vec{q}) \phi(\Omega,\vec{q})$, where $\Omega$ is the frequency of field $\phi(t,\vec{r})$. We use the Keldysh method for non-equilibrium systems \cite{Keldysh} to find the generalized Kubo formula:
\begin{eqnarray}\label{resp1}
\chi_n(\Omega,\vec{q})=i\frac{\omega}{2\pi}\int_0^{2\pi/\omega}e^{in\omega T} dT\int_0^\infty dt\nonumber\\  \langle\left[\hat{\rho}(\vec{q},T-t/2)\ \hat{\rho}(-\vec{q},T+t/2)\right]_- \rangle e^{i\Omega t}.
\end{eqnarray}
where the brackets stand for the matrix element over the driven state with electrons filling the Fermi circle of radius $p_F$ and the density operators $\hat{\rho}$ are given by Eq.(\ref{Density}). After taking the integral with respect to $T$ and $t$ we find:
\begin{eqnarray}\label{Responce} 
\chi_n(\Omega,\vec{q}) = \frac{m}{2\pi}\sum_{k=-\infty}^{\infty} J_k\left(\frac{1}{\omega}\vec{V}\vec{q} \right) J_{k+n}\left(\frac{1}{\omega}\vec{V}\vec{q} \right) \nonumber\\ \left(1 - i\frac{P(\Omega+q^2/2m)-P(\Omega-q^2/2m)}{q^2/m}\right),
\end{eqnarray}
where $P(x)=\sqrt{v_F^2q^2-(x+(k+n/2)\omega)^2}$. The static response function for $\omega\ll v_Fq\ll \epsilon_F$ is found as:
\begin{equation} \label{chi}
\chi_{2n}(\vec{q})= \frac{m}{2\pi}J_n^2 \left( \vec{V}\vec{q}/\omega\right).
\end{equation}
Zeroes of this response is an analog of the phenomenon of the coherent destruction of tunneling \cite{gd91}.

\section{Quasistatic approximation} 

In this section we introduce a quasistatic approximation and we discuss the role of disorder and electron-phonon interaction.

For slow electric fields it is convenient to take the Fourier transform of Eq.(\ref{resp}) with respect to the 'fast' time $t-t'$: $\langle \hat{\rho}(\Omega,\vec{q},t)\rangle =\chi(\Omega,\vec{q},t) \phi(\Omega,\vec{q})$. For strong electric fields: $\omega\ll p_FV$, a large number of harmonics contributes to the sum (\ref{Responce}). In this limit the response function (\ref{Responce}) has an asymptote:
\begin{equation}\label{Responce1}
\chi(\Omega,\vec{q},t) = \frac{m}{2\pi}\left(1 - i\frac{P(\Omega+\frac{q^2}{2m})- P(\Omega-\frac{q^2}{2m})} {q^2/m}\right),
\end{equation}
where $P(x)=\sqrt{v_F^2q^2-\left(x+\vec{V}(t)\vec{q}\right)^2}$. Eqs.(\ref{Responce},\ref{Responce1}) become identical in the limit $\omega\ll p_FV\ll\epsilon_F$ and $q\ll p_F$. 

The Green functions in the laboratory and in the center of mass frames are defined as: $G_{lab}(t,t',\vec{p})=-i\langle T\hat{\psi}(\vec{p},t)\hat{\psi}^\dagger(\vec{p},t')\rangle$ and
\begin{equation}\label{CMGreen}
G(t,t',\vec{p})=-i\langle T\hat{\chi}(\vec{p},t)\hat{\chi}^\dagger(\vec{p},t')\rangle,
\end{equation}
where $T$ is the chronological ordering operator. These two Green functions are related by the canonical transformation (\ref{Canonical}). The Green function in the center of mass frame (\ref{CMGreen}) depends on $t-t'$ and is determined by the Hamiltonian $\hat{H}_0+\hat{H}^{int}$ of the static Fermi liquid:
\begin{equation}
G(\epsilon,\vec{p})=\frac{a}{\epsilon-\xi(p)+i\eta (p-p_F)|p-p_F|},
\end{equation}
where $a$, $\eta$ are the amplitudes of the pole renormalization and relaxation \cite{AGD}. In the Fermi gas $a=1$ and $\eta \rightarrow +0$. 

\textbf{Disordered Fermi gas} (non-interacting) can be described by the model of weak short ranged impurity potential treated in the Born approximation. The disorder Hamiltonian reads: $\hat{H}^{imp}= \int\ U(\vec{r}) \hat{\rho}(t,\vec{r})\ d^2\vec{r}$, where the random Gaussian disorder potential has the correlation function: $\langle U(\vec{0})U(\vec{r}) \rangle= \delta(\vec{r})/ 2\pi\nu_F\tau$. $\tau$ is the mean elastic scattering time and $\nu_F=m/2\pi$ is the 2D density of states. Disorder breaks down the translational symmetry and matrix elements of the disorder Hamiltonian become time dependent in the center of mass frame. On the other hand the Green function averaged over the disorder is translationally invariant \cite{AGD} both in the static and driven Fermi gases. In the standard diagrammatic approach \cite{AGD} that neglects diagrams with crossing impurity lines (small as $1/\epsilon_F\tau$) and in the limit $\omega\ll |t-t'|^{-1}\ll \epsilon_F$ we find the electron self-energy:
\begin{equation} \label{ImpSE}
\Sigma(t-t',\vec{p})=\int e^{i\vec{V}(t)(\vec{p'}-\vec{p})(t-t')} \frac{G(t-t',\vec{p'})}{2\pi\nu_F\tau} \frac{d^2\vec{p'}}{(2\pi)^2},
\end{equation}
depending on the time $t$ as a parameter. Taking the Fourier transform of Eq.(\ref{ImpSE}) with respect to $t-t'$ and using the definition of the self-energy: $G^{-1}(\epsilon)= \epsilon-\xi(p)- \Sigma(\epsilon,\vec{p})$, where $\xi(p)=p^2/2m-\mu$, we find:
\begin{equation} \label{ImpSigma}
\Sigma(\epsilon,\vec{p})=\frac{1}{2\pi\tau}\int \frac{1}{\epsilon_+-\xi'-\Sigma(\epsilon_+,\vec{p'})}\ d\xi' ,
\end{equation}
where $\epsilon_+=\epsilon+\vec{V}(t)(\vec{p'}-\vec{p})$. The solution of the Dyson equation (\ref{ImpSigma}) yields: $\Sigma(\epsilon,t,\vec{p})= -i\,\textrm{sgn}(\epsilon- \vec{V}(t)\vec{p})/2\tau$ (neglecting the time dependent real part) and the Green function in the center of mass frame:
\begin{equation}\label{ImpGreen}
G(\epsilon,\vec{p})=\frac{1}{\epsilon-\xi(p)+i\ \textrm{sgn}\left(\epsilon- \vec{V}(t)\vec{p}\right) /2\tau}.
\end{equation}
The corresponding Green function in the laboratory frame is: $G_{lab}(\epsilon,\vec{p})= (\epsilon+ \vec{V}(t)\vec{p}- \xi(p)+i/2\tau\ \textrm{sgn}(\epsilon))^{-1}$

One Green function (\ref{ImpGreen}) describes two distinct situations. In the absence of electron-phonon interaction the electron distribution function $f(\vec{p})=1$ if $\epsilon- \vec{V}(t) \vec{p}<0$ and zero otherwise. In the center of mass frame this Fermi circle oscillates slowly. On the other hand when electrons could quickly lose the energy into the bulk the driven state is incoherent with $f(\vec{p})=1$ inside the stationary Fermi circle. Here scattering between few electron states close to the Fermi circle with the reversed imaginary parts: $\epsilon- \vec{V}(t)\vec{p}<0$ and $p>p_F$ or $\epsilon- \vec{V}(t)\vec{p}>0$ and $p<p_F$, creates the Joule heating.  

\textbf{Phonon cooling} is conveniently described by the electron-phonon Hamiltonian in the deformation potential approximation (Froelich model):
\begin{equation}\label{Froelich}
\hat{H}^{e-ph}=\Theta\int\ \hat{\rho}(\vec{r},t)\ \textrm{div} \vec{u}(\vec{r},t)|_{z=0}\ d^2\vec{r},
\end{equation}
where $\Theta$ is the isotropic deformation potential and $\hat{\rho}(\vec{r},t)$ is given by Eq.(\ref{Density}). Electron self-energy is found in the second order of the electron-phonon interaction (\ref{Froelich}):
\begin{equation}\label{PhonSE}
\Sigma(\epsilon,\vec{p})=i\frac{\Theta^2}{\rho s^2}\int G(\epsilon+\omega,\vec{p'}) D\left(\tilde{\omega},\vec{q}\right) \frac{d\omega}{2\pi}\frac{d^2\vec{p'}dq_z}{(2\pi)^3},
\end{equation}
where $\rho$ is the crystal density, $\tilde{\omega}=\omega+\vec{V}(t)\vec{q}$, $\vec{q}$ is the 3D phonon wave vector. There is in-plane momentum conservation: $\vec{q}_{||}=\vec{p'}-\vec{p}$. The velocity of free electrons in \textit{ac}-field is $V(t)$. The phonon propagation for zero lattice temperature is retarded:
\begin{equation}\label{PhonGreen}
D(\omega,\vec{q})=\frac{(sq)^2}{\omega^2-(sq)^2+i0}
\end{equation} 
We transform the measure of integration in Eq.(\ref{PhonSE}). Let $\phi(q)$ be the angle between the momentum $\vec{p'}$ and the momentum $\vec{p}$. It can be found from the equation: $p^2+p'^2- 2pp'\cos{\phi(q)} =q^2_{||}$. We substitute $\phi(q)$ in the measure of integration $d^2\vec{p'}= p'dp'd\phi(q)$  and find the transformed measure: $q_{||}p'dp'dq_{||}/\sqrt{p_F^2q_{||}^2-q_{||}^4/4-(p^2-p'^2)^2/4}$. In the limit $p^2-p'^2\ll p_Fq_{||}$  Eq.(\ref{PhonSE}) can be rewritten as
\begin{equation}\label{PhonSE1}
\Sigma(\epsilon)=i\frac{\nu_F s^3}{E_{ph}^2}\int \frac{G(\epsilon+\omega,\xi')D(\tilde{\omega},\vec{q})} {p_F\sqrt{1-q_{||}^2/4p_F^2}} \frac{d\omega dq_{||} dq_z d\xi'}{(2\pi)^2} 
\end{equation}
where $E_{ph}=\sqrt{\rho s^5/\Theta^2}$ is the characteristic phonon energy. Imaginary part of the self-energy gives the probability of the \textit{ac}-field assisted phonon emission. Multiplying it by the energy transfer $\vec{V}(t)\vec{q}_{||}-sq$ we find the cooling rate. Electron-electron interaction establishes the Fermi-Dirac electron distribution function: $f(\epsilon/T^*)= 1/(\exp(\epsilon/T^*)+1)$ with the electron temperature $T^*$ despite the bulk being kept at zero temperature. 

Following the standard method \cite{AGD} we find the following results. As long as $V(t)<s$ there is no energy absorption from the \textit{ac}-field into the electron system associated with the phonon emission. Using the identity:  
\begin{equation}
\int_{-\infty}^\infty f\left(\frac{\epsilon}{T}\right) \left[1-f\left(\frac{\epsilon -\omega}{T}\right)\right] d\epsilon= \omega N\left(\frac{\omega}{T}\right),
\end{equation}
where $N(\omega/T)=1/(\exp(\omega/T)-1)$ is the Bose-Einstein distribution function, we find cooling rate per unit area:
\begin{equation}
\frac{dE}{dt}=- \frac{\nu_F^2s^4}{2p_FE_{ph}^2}\int (\delta s\ q)^2N\left(\frac{\delta s\ q}{T^*}\right)  \frac{q^2dqd\theta d\phi} {(2\pi)^3},
\end{equation}
where $\delta s=s-V(t)\cos(\theta)\cos(\phi)$, $q_z=q\sin\theta$, $q_{||}=q\cos\theta$ and $\phi$ is the angle between vectors $\vec{V}(t)$ and $\vec{q}_{||}$. Evaluating the integral we find the cooling rate:
\begin{equation}\label{Cooling}
\frac{dE}{dt}=-\frac{12\zeta(5)s\nu_F^2 T_*^5}{2\pi^2 p_F E_{ph}^2} \left(\frac{\textbf{K} (k)}{1-k^2}+ \frac{(3-k^2) \textbf{E}(k)} {\left(1-k^2\right)^2}\right),
\end{equation}
where $k=V(t)/s$. The cooling rate diverges as $V(t)\rightarrow s$.

At $V(t)>s$ a channel for heating is opened. Electrons in a narrow area in momentum space emit phonons and absorb the $ac$-field energy. This part of the heating rate:
\begin{equation}\label{Heating}
\frac{dE}{dt}=\frac{64 p_F^2}{90\pi E_{ph}^2}\left(V(t)-s\right)^4 p_F^4 \theta\left(V(t)-s\right),
\end{equation}  
where $\theta(x)$ is the step-function, is additive to the cooling rate (\ref{Cooling}) from the rest of electron states. Heating (\ref{Heating}) does not depend on $T^*$ as long as it is small. And the balance of energy flows Eqs.(\ref{Cooling},\ref{Heating}) makes the electron temperature to grow rapidly: $T^*\sim |V(t)-s|^{6/5} v_F^{2/5}s^{-3/5}p_F$. Thus there is a kinetic transition from the fast cooling Eq.(\ref{Cooling}) to the growing heating Eq.(\ref{Heating}) at the critical value of \textit{ac}-field: $V(t)=s$. 

\section{Electron in crossed magnetic and \textit{ac}-electric fields}

In this section we find the wave function, density and current operators of free electrons in crossed magnetic and periodic \textit{ac}-electric fields.

Consider an electron confined to a plane in magnetic field $H$ perpendicular to the plane and driven by in-plane electric field $E(t)=E\cos(\omega t)$ that is linearly polarized in $x$ direction. In Landau gauge: $A_x=-Hy$ , and in the vector potential gauge for the uniform \textit{ac}-field: $A_x(t)=-cE\sin(\omega t)/\omega$, the Hamiltonian reads:
\begin{equation}\label{HamPC}
\hat{H}_0=\frac{1}{2m}\left(-i\frac{\partial}{\partial x}+\frac{e}{c}(-Hy+A_x(t))\right)^2- \frac{1}{2m} \frac{\partial^2}{\partial y^2}
\end{equation}
In high Landau level $N$ there are two distinct lengths: the cyclotron radius $R_c= \sqrt{2N} l_H$ and the magnetic length $l_H=\sqrt{\hbar c/eH}$. We use the magnetic units where $e=c$, $H=1$ and $l_H=1$. A characteristic energy of the \textit{ac}-field is defined as: $\mathcal{E}=eER_c$. The electron wave function in Landau level $N$ and in the state of the gauge index $p$: \cite{ryzhii}
\begin{eqnarray} \label{RyzhWF}
\psi_{Np}(t,\vec{r})=\frac{\exp(ipx)}{\sqrt{\sqrt{\pi}L_x 2^N N!}} H_N\left( y-p-Y(t) \right)  \nonumber\\ \exp\left({iX(t)(y-p-Y(t))-(y-p-Y(t))^2/2+i\Phi}\right)
\end{eqnarray}
is the solution of the time dependent Schroedinger equation with the Hamiltonian (\ref{HamPC}). $X(t)$ and $Y(t)$ are the $x,y$-coordinates of the electron classical trajectory - an ellipsoid - in the crossed electric and magnetic fields:
\begin{equation} \label{ellips}
X(t)=\frac{{\cal E}\omega_H\cos(\omega t)}{\sqrt{2N}\omega^2_-}, \ 
Y(t)=\frac{{\cal E}\omega^2_H\sin(\omega t)}{\sqrt{2N}\omega\omega^2_-},
\end{equation}
where $\omega_H=eH/mc$, $\omega_\pm^2=\omega^2\pm\omega_H^2$ and: 
\begin{equation}
\Phi=\frac{{\cal E}^2\omega_H\omega^2_+} {16N\omega\omega^4_-} \sin(2\omega t) -\omega_Ht \left(N+\frac{1}{2}+\frac{{\cal E}^2}{8N\omega^2_-} \right)
\end{equation}
is the wave function phase. The retarded bare electron Green function is diagonal in the indices $N$ and $p$: $g_N^{R}(\epsilon)=1/(\epsilon-N\omega_H +i\delta)$, where $0<\delta\rightarrow 0$. 

Using the wave function (\ref{RyzhWF}) we find the density operator $\hat{\rho}(t,\vec{r})=  \hat{\psi}^\dagger(t,\vec{r}) \hat{\psi}(t,\vec{r})$ of the driven 2DEG as:
\begin{equation} \label{rho_mag}
\hat{\rho}(t,\vec{q})=\hat{\rho}_0(\vec{q})\exp\left(-iq_y X(t)+ iq_x Y(t)\right),
\end{equation}
where the density operator of the stationary 2DEG reads:
\begin{equation} \label{rho0}
\hat{\rho}_0(\vec{q})=\sum_{nn'}V_{nn'}(\vec{q}) \sum_p \psi_{n'p}^\dagger \psi_{n\ p+q_x} e^{iq_y(p+q_x/2)-q^2/4}.
\end{equation}
The vertex $V_{nn'}(\vec{q})$ for electron scattering from the Landau level $n$ to the Landau level $n'$ with transferred momentum $\vec{q}$ is defined as:
\begin{equation}\label{VertexLaguerre}
V_{nn'}(\vec{q})=\left(\frac{n'!}{n!}\right)^{\pm 1/2} \left(\frac{q_y\pm iq_x}{\sqrt{2}}\right)^{|n-n'|}   L^{|n-n'|}_{{\rm min}(nn')} \left(\frac{q^2}{2}\right)
\end{equation}
where $L_m^k(x)$ is the Laguerre polynomial. The two signs $\pm$ in Eq.(\ref{VertexLaguerre}) corresponds to the two cases $n>n'$ and $n<n'$.

Instantaneous electron-electron interaction is invariant under the transformation (\ref{rho_mag}) of the density operators from the stationary to the driven form for any time dependence of the \textit{ac}-field. For instance the Laughlin wave function is as good for driven 2DEG as for the stationary one. Variation of the Hamiltonian (\ref{HamPC}) with respect to the vector potential $\delta \vec{A}(\vec{r})$ gives the current operator:
\begin{eqnarray} \label{currentOp}
\hat{J}(t,\vec{r})=\frac{1}{\sqrt{2}}\left(\frac{\partial}{\partial y}+ i\frac{\partial}{\partial x}+A_x(t)+y\right)= \nonumber\\ =\hat{\Pi}+(i\omega_H\cos(\omega t)+ \omega\sin(\omega t) )\ \mathcal{E}/2\sqrt{N}\omega^2_-
\end{eqnarray}
where the operator $\hat{\Pi}$ lowers the Landau level state $\Pi|N,p\rangle =\sqrt{N} |N-1,p\rangle$. $\hat{J}^\dagger(t,\vec{r})$ is a Hermitean conjugated operator to $\hat{J}(t,\vec{r})$. In Sec. V we neglect the second term in the r.h.s. of Eq.(\ref{currentOp}) as being small as $1/N^2$ in the high Landau levels. We will study effects of \textit{ac}-field in the leading order of $1/N^0$.

\section{Disordered 2DEG in the quantum limit}

In perpendicular magnetic field that quantizes the electron motion, electron states are degenerate and it is the disorder that lifts this degeneracy and imparts a velocity to electrons. Because disorder couples to \textit{ac}-field the problem of Landau level broadening and \textit{dc}-conductivity has to be reexamined. In this section we solve this problem in high non-overlapping Landau levels and for short-range correlated Gaussian disorder. 

The Boltzmann kinetic equation is useless for calculating the conductivity in the quantum limit. The current is the sum: $\langle \vec{j}\rangle= e\sum_i \vec{v}_i\delta f_i$, over electron states $i$ and a variation of the electron distribution function in weak $dc$-electric field is $\delta f_i\sim \vec{v}_i\vec{E}$. Because $\vec{v}_i$ is random in the quantum limit it follows that  $\langle\vec{v} \rangle=0$ and $\langle \delta f\rangle =0$. Instead we adopt a diagrammatic approach that calculates an average of the diffusion coefficient $D_i\sim \vec{v}^2_i$. 

We assume that the transport of 2DEG can be properly described in terms of the non-interacting electrons. In the framework of Landau theory Eliashberg has proved that the conductivity of Fermi liquid is the conductivity of non-interacting quasiparticles near the Fermi surface \cite{Elias}. In magnetic fields the de Haas-van Alphen oscillations of the magnetic moment has also been described by the non-interacting quasiparticles \cite{bg61}. 

The impurity scattering dominates the transport at low temperature. The Hamiltonian of electron moving in a random impurity potential $U(\vec{r})$ reads:
\begin{equation}\label{Hamiltonian2DEG}
\hat{H}^{imp}=\int\psi^\dagger(\vec{r})U(\vec{r})\psi(\vec{r}) d^2\vec{r},
\end{equation}
where we omit the spin index of the creation and annihilation operators: $\psi^\dagger$ and $\psi$ due to the vanishing exchange interaction in high Landau levels. The probability distribution functional of the impurity potential $U(\vec{r})$ is assumed to be Gaussian with the correlation function: 
\begin{equation}\label{UU}
\langle U(\vec{\vec{r}})U(\vec{r'})\rangle =2\pi ul_H^2\delta\left(\vec{r}-\vec{r'}\right), \ \text{where} \ u=\omega_H/\tau,
\end{equation}
where $\tau$ is the mean elastic scattering time. 

We expand the electron Green function:
\begin{equation}\label{GreenGen}
G(t,t';\vec{r},\vec{r'})=-i\langle T\hat{\psi}(t,\vec{r})\hat{\psi}^\dagger(t',\vec{r'})\rangle,
\end{equation}
into a series over the impurity potential $U(\vec{r})$ and then average it over the disorder, using the standard diagrammatic method of crossed diagrams \cite{AGD}. In Landau level the coordinate dependence of the average Green function $G(t,t';\vec{r},\vec{r'})= G_N(t,t')\Phi_N(\vec{r},\vec{r'})$ is given by the universal phase factor $\Phi_N(\vec{r},\vec{r'})$ which is conveniently transfered onto the vertex elements Eq.(\ref{rho0},\ref{VertexLaguerre}). $G_N(t,t')$ is the sum of diagrams consisting of bare electron $g_N(\epsilon)$ and impurity (\ref{UU}) lines incident in vertices (see Eq.(\ref{VertexLaguerre})): $V_{nn'}(\vec{q}) \exp(iq_x (p+q_y/2)-q^2/4) \delta_{p',p+q_y}$, that describe a process of electron scattering off the impurity potential $U(\vec{q})$ from the state $(n,p)$ into a state $(n',p')$ with transfered momentum $\vec{q}$. 

In the absence of \textit{ac}-field the density of states in a high Landau level follows the semi-circle law: $\rho(\epsilon)= \sqrt{4u-\epsilon^2}/2\pi l_H^2 u$, where $\epsilon$ is zero in the center of the Landau level. At $T=0$ the electron distribution function $f(\epsilon)$ is given by the step function with the Fermi energy being determined by the electron density $n=(N+\nu)/2\pi l_H^2$, where $\nu$ is the filling factor of the last $N$th Landau level. The longitudinal conductivity: $\sigma_{xx}=\pi e^2Nu\rho^2(\epsilon_F)$,\cite{kubo65} is comparable to the Hall conductivity $\sigma_{xy}=ecn/H$. 

Consider 2DEG driven by strong and slow linearly-polarized \textit{ac}-electric field: $\vec{\mathcal{E}}= \vec{e}_x\mathcal{E} \cos(\omega t)$, where $\sqrt{u}\ll \mathcal{E}\ll\omega_H$ and $\omega\ll u/\mathcal{E}$. We neglect Landau level mixing small as $\sqrt{u}/\omega_H$ and impurity lines crossings small as $1/N$ and omit the index $N$ from the notations. The self-energy is given by the sum of all non-crossing diagrams:
\begin{equation} \label{SE}
\Sigma(t,t')=u(t,t')G(t,t'),\ \Sigma(\epsilon,t)=\!\int\! \frac{d\Omega}{2\pi} u(\Omega,t) G(\epsilon+\Omega,t)
\end{equation}
The impurity line includes the correlation function (\ref{UU}) and the two adjacent vertices Eq.(\ref{rho_mag},\ref{rho0},\ref{VertexLaguerre}):
\begin{equation} \label{ImpLineTime}
u(t,t')=u\int \left|V_{NN}(\vec{q})\right|^2 \ e^{i\vec{q}\delta\vec{r}(t,t')-q^2/2}\  \frac{d^2\vec{q}}{(2\pi)^2}.
\end{equation}
where $\delta \vec{r}(tt')=\vec{R}(t)-\vec{R}(t')$ is the differential along the classical electron trajectory (\ref{ellips}). The integration with respect to the transferred momentum $\vec{q}$ gives:
\begin{equation} \label{ImpLine}
u(t,t')=u\left[L_N^0\left(\delta r^2(tt')/2\right)\right]^2\exp\left(-\delta r^2(tt')/2\right),
\end{equation} 
where $L_n^m(x)$ is the Laguerre polynome. In the limit: $N\gg 1$ and $\omega|t-t'|\ll 1$, Eq.(\ref{ImpLine}) has an asymptote:
\begin{equation} \label{ImpLine1}
u(t,t')=uJ^2_0\left(\mathcal{E}c(t)(t-t')\right),
\end{equation}
where $c(t)=\sqrt{\cos^2\omega t+(\omega/\omega_H)^2\sin^2\omega t}$. The Fourier transform of the impurity line with respect to the time $t-t'$ reads:
\begin{equation}\label{FourierU}
u(\Omega,t)=2\Delta(t)\textbf{K}\left(\sqrt{1-\frac{\Omega^2}{4 \mathcal{E}^2 c^2(t)}}\right),
\end{equation}
where $\Delta(t)=u/\pi\mathcal{E}c(t)$ is the characteristic broadening of the Landau level, ${\bf K}(k)$ is the Jackobi elliptic function of the first kind. $u(\Omega,t)=0$ for $|\Omega|> 2\mathcal{E}c(t)$. The non-crossing approximation is valid provided $\delta r(tt')\ll l_H$ or $\mathcal{E}c(t) |t-t'|\ll \sqrt{N}$. In the limit $1\ll\mathcal{E}c(t)|t-t'|$ we find approximately: $u(t,t')= \Delta(t)/|t-t'|$, where oscillations with the frequency $\epsilon_{up}=2\mathcal{E}c(t)$ are neglected as resulting in small corrections of the order of $\sqrt{u}/\mathcal{E}c(t)$.

For a free electron in random potential the Green function $G(t,t')$ can be chosen retarded with the property $G^R(t,t')=0$ at $t<t'$. Therefore we substitute $|t-t'|=t-t'$ in $u(t,t')= \Delta(t)/|t-t'|$ and take the Fourier transform of the self-energy (\ref{SE}) with respect to the 'fast' time $t-t'$:
\begin{equation} \label{Dyson}
\frac{d\Sigma^R}{d\epsilon}=i\Delta(t) G^R(\epsilon),
\end{equation}
where the 'slow' time $t$ serves as a parameter and we omit it from the notations. Solution of the equation $G^{-1}(\epsilon)= \epsilon-\Sigma(\epsilon)$ together with the Dyson equation (\ref{Dyson}) yields the Green function having an asymptote: $G^R(\epsilon)=1/\epsilon$ as $|\epsilon|\rightarrow\infty$, implicitly as:
\begin{equation} \label{AvG}
\epsilon=\frac{1}{G^R(\epsilon)}-i\Delta(t)\log\left(\frac{\mathcal{E}c(t)}{\Delta(t)-iG^{-1}_R(\epsilon)}\right).
\end{equation}
For small energies $\epsilon\sim\Delta(t)\ll \mathcal{E}c(t)$ we find $G_R^{-1}(\epsilon)\approx \epsilon+ i\Delta(t) \log\left(\mathcal{E}c(t)/\Delta(t)\right)$. Advanced Green function is complex conjugated to the retarded one: $G^A(\epsilon)=(G^R(\epsilon))^*$.

In driven 2DEG $\rho(\epsilon)=\textrm{Im} G^A(\epsilon)/\pi$ is the spectral density of electron states - an analog of the density of states in the stationary 2DEG. The spectral density (\ref{AvG}) has an asymptote: $\textrm{Im} G^A(\epsilon)= \Delta(t)\log\left(\mathcal{E}c(t)/\epsilon\right)/\epsilon^2$, as $\Delta(t) \ll |\epsilon|\ll \mathcal{E}c(t)$. At small $\epsilon$ the spectral density has approximately the Lorentz form: $\textrm{Im} G^A(\epsilon)=\Delta L/(\epsilon^2+\Delta^2L^2)$, where $L= \log\left(\mathcal{E}c(t)/ \Delta(t)\right)$. At high frequencies $|\epsilon|> 2\mathcal{E}c(t)$ we find a crossover tail $\textrm{Im}G^A(\epsilon)\sim \Delta^2/\epsilon^2(|\epsilon|-2\mathcal{E}c(t))$ towards the self-consistent Born approximation solution $\rho(\epsilon)=0$ for $|\epsilon|>2\sqrt{u}$. Using Eq.(\ref{Dyson}) we prove that the spectral density is normalized:
\begin{equation}
-\int \textrm{Im} G^R(\epsilon)\ d\epsilon =\pi
\end{equation}
We substitute $G^R d\epsilon= dG_R^{-1}/(G_R^{-1}+i\Delta)$ into the integrand $2i\textrm{Im} G^R d\epsilon=G^R d\epsilon-G^A d\epsilon$ and deform the contour of integration to the real axis: $\int_C\ dG_R^{-1}/(G_R^{-1}+i\Delta)- \int_C\ dG_A^{-1}/(G_A^{-1}-i\Delta)=-2i\int dG\Delta/(G^2+\Delta^2)=-2i\pi$.

Retarded or advanced Green function alone could not describe many electron systems and the electron distribution function $F(\epsilon)$ has to be determined as well. We derive $F(\epsilon)$ from the average Keldysh Green function \cite{Keldysh}. We sum up non-crossing diagrams using the Keldysh method \cite{Keldysh} and find the Dyson equation for $F(\epsilon)$:
\begin{eqnarray}\label{FDyson}
\left[F(\epsilon)-f(\epsilon)\right]\textrm{Im} G^R(\epsilon)= G^R(\epsilon)G^A(\epsilon) \int\frac{d\Omega}{2\pi} \nonumber\\ u(\Omega) \left[F(\epsilon+\Omega)-f(\epsilon)\right] \textrm{Im} G^R(\epsilon+\Omega), 
\end{eqnarray}
where Green functions are given by Eq.(\ref{AvG}). At $T=0$ the initial electron distribution function $f(\epsilon)$ is the Fermi-Dirac step function. We prove the identity 
\begin{equation}\label{WardId}
G^R(\epsilon)G^A(\epsilon) \int\frac{d\Omega}{2\pi} u(\Omega) \textrm{Im} G^R(\epsilon+\Omega) =\textrm{Im} G^R(\epsilon),
\end{equation}
by evaluating the imaginary parts of the two equations: i) $G(\epsilon)=1/(\epsilon-\Sigma(\epsilon))$ and ii) the Dyson Eq.(\ref{SE}). From i) we find that $\textrm{Im} G(\epsilon)= G^R(\epsilon)G^A(\epsilon) \textrm{Im}\Sigma(\epsilon)$. Then from the Dyson equation (\ref{SE}) the identity (\ref{WardId}) follows. A consequence of this identity is that the solution of the Dyson equation (\ref{FDyson}) does not depend on the initial distribution function $f(\epsilon)$. Solving Eq.(\ref{FDyson}) for $F(\epsilon)$ and using the normalization of the total number of electrons we find that $F(\epsilon)= \nu$, where $\nu$ is the filling factor of the $N$th Landau level. This solution interpolates between $F=1$ and $F=0$ for $N-1$ and $N+1$ Landau levels making connection in the gap regions where $\rho(\epsilon)=0$. The constant distribution function is an analog of the classical energy diffusion for the non-stationary scattering.\cite{Fermi} 

\begin{figure}
\includegraphics{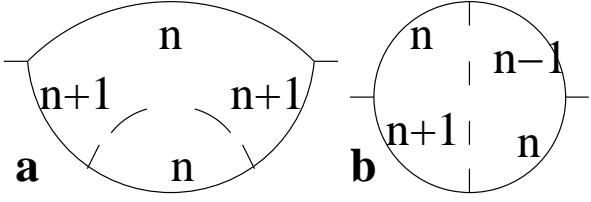}
\caption{Two diagrams: a and b, that give the impurity conductivity in strong magnetic fields}
\end{figure}

We use the Keldysh method \cite{Keldysh} to find the diagonal conductivity. It is explicitly a real function: $\sigma_{xx}(\Omega)= \sigma_{xx}(-\Omega)^*$, with the reactive and the inductive parts: $Z(\Omega)= R(\Omega^2)+i\Omega L(\Omega^2)$, and we disregard the inductive part here. For the reactive part we write:
\begin{equation}\label{SigmaKG}
\sigma_{xx}(\Omega)=e^2\omega_H^2\int_{-\infty}^{\infty}\!\frac{d\epsilon}{2\pi}\  \frac{F(\epsilon)-F(\epsilon-\Omega)}{\Omega} \sigma_{xx}(\Omega,\epsilon),
\end{equation}
where the conductivity of the states with energy $\epsilon$ is:
\begin{eqnarray}\label{Sigma}
\sigma_{xx}(\Omega,\epsilon)+\sigma_{yy}(\Omega,\epsilon)= 2\sum_{nn'=1}^\infty\! \sqrt{n'n}\ \ \langle \textrm{Im} G_{nn'}(\epsilon-\Omega) \nonumber\\ \textrm{Im} G_{n-1,n'-1}(\epsilon)+\textrm{Im} G_{nn'}(\epsilon) \textrm{Im} G_{n-1,n'-1}(\epsilon-\Omega)  \rangle ,
\end{eqnarray}
that includes inter Landau level transitions. The disorder average $\langle ... \rangle$ is expanded into a series of diagrams. In the leading order the conductivity is the sum of two diagrams in Fig.1 (where $\omega$ is no longer the \textit{ac}-frequency):
\begin{eqnarray}
\sigma_{xx}(\Omega)\pm\sigma_{yy}(\Omega)=2e^2N\int\!\!\int \textrm{Im} G(\epsilon+ \omega-\Omega) \nonumber\\ \textrm{Im} G(\epsilon) \left\{ \begin{array}{c} P(t,\omega) \\ S(t,\omega) \end{array} \right\} \frac{F(\epsilon+\omega-\Omega)- F(\epsilon)}{\Omega} \frac{d\epsilon d\omega}{(2\pi)^2}.
\end{eqnarray}
Functions $P(t,\omega)$ and $S(t,\omega)$ are given by the sum of the two impurity lines on diagrams a) and b) in Fig.1. $P(t,\omega)$ corresponds to diagrams with the two different $J$ and $J^\dagger$ vertices (shown in Fig.1) whereas $S(t,\omega)$ corresponds to the diagrams with $J-J$ or $J^\dagger-J^\dagger$ pair of vertices (not shown): $P(t,t-t')=uJ_0^2\left(x\right)- uJ_1^2\left(x\right)$ and $S(t,t-t')=uJ_0\left(x\right)J_2\left(x\right) +uJ_1^2\left(x\right)$, where $x=\mathcal{E}c(t)(t-t')$. Thus, the diagonal conductivity is anisotropic with the two axis being defined by the linearly polarized \textit{ac}-field. $P(t,\omega)$ and $S(t,\omega)$ are non-zero if $|\omega|<2{\cal E}c(t)$:
\begin{eqnarray}
&& P(t,\omega)=\frac{4u}{\pi\mathcal{E}c(t)} \textbf{E}\left(k\right), \nonumber\\ && S(t,\omega)=\frac{2u}{\pi\mathcal{E}c(t)}\left[ \frac{\omega^2}{\mathcal{E}^2 c^2(t)} \textbf{K}\left( k\right)- 2\textbf{E}\left(k\right) \right],
\end{eqnarray}
where ${\bf E}(k)$ is the Jackobi elliptic function of the second kind and $k=\sqrt{1- \omega^2/4 \mathcal{E}^2 c^2(t)}$ is the elliptic modulus. As $P(t,\omega)=P(t,-\omega)$ and $S(t,\omega)=S(t,-\omega)$ are even functions we work out the \textit{dc}-conductivity:
\begin{eqnarray} \label{Conductivity}
\sigma_{xx}\pm\sigma_{yy}=2e^2N\int \frac{d\omega d\epsilon}{(2\pi)^2} \left\{ \begin{array}{c} P(t,\omega) \\ S(t,\omega) \end{array} \right\} \textrm{Im} G(\epsilon) \nonumber\\ \left[(F(\epsilon)-F(\epsilon+\omega))\frac{d}{d\epsilon}\textrm{Im} G(\epsilon+\omega) - \textrm{Im} G(\epsilon+\omega) \frac{dF}{d\epsilon} \right]
\end{eqnarray}
In the limit $\omega\ll \mathcal{E}c(t)$ the parallel to the \textit{ac}-field conductivity $\sigma_{xx}$ is by the factor $\Delta^2(t)/\mathcal{E}^2c^2(t)$ smaller than the perpendicular to \textit{ac}-field conductivity $\sigma_{yy}$. This corresponds to the geometry of the electron drift. In the absence of the electron-phonon interaction and the Landau level mixing $F(\epsilon)=\nu$ and according to Eq.(\ref{Conductivity}) both components of the diagonal \textit{dc}-conductivity are zero. 

Electron-phonon scattering invalidates the $F(\epsilon)=\nu$ solution. For $\epsilon_F \mathcal{E}c(t) \ll E_{ph}^2$, where $E^2_{ph}=\rho s^5/\Theta^2$ is the characteristic phonon energy, the electron-electron scattering is dominant. In this case the electron distribution function is the Fermi-Dirac one: $F(\epsilon)= 1/(\exp(\epsilon-\mu)/T^*)+1)$, where the chemical potential $\mu$ is determined from the total electron normalization condition whereas the effective electron temperature $T^*$ is found from the balancing of disorder and electron-phonon scattering. A simple analysis shows that electron distribution function is spread over many Landau levels in this case. On the other hand in the limit $E_{ph}^2 \ll \epsilon_F \mathcal{E}c(t)$ the electron-electron interaction can be neglected and weak electron-phonon interaction can be treated in the second order of the perturbation theory. In the limit $V(t)\ll s$ the collision integral balancing of disorder and phonon scattering processes reads:
\begin{eqnarray} \label{Collision}
&&\textrm{Im} G(\epsilon)\int\frac{d\omega}{2\pi} u(\omega) \left[F(\epsilon+\omega)-F(\epsilon)\right] \textrm{Im} G(\epsilon+\omega)= \nonumber\\ && \textrm{Im} G(\epsilon)\int\frac{s^4q}{2E_{e-ph}^2}\left\{ F(\epsilon+sq)\left[1-F(\epsilon)\right]\textrm{Im} G(\epsilon+sq) \right.\nonumber\\ && \left. - F(\epsilon)\left[1-F(\epsilon-sq)\right]\textrm{Im} G(\epsilon-sq) \right\}\frac{d^3\vec{q}}{(2\pi)^3}.
\end{eqnarray}  
It can be linearized by an expansion $F(\epsilon)=\nu+\delta F(\epsilon)$:
\begin{eqnarray}\label{deltaF}
\int\frac{d\omega}{2\pi} u(\omega) \left[\delta F(\epsilon+\omega)-\delta F(\epsilon)\right] \textrm{Im} G(\epsilon+\omega)= \nonumber\\ \int\frac{d^3\vec{q}}{(2\pi)^3}\ \frac{\nu(1-\nu)\ s^4q}{2E_{ph}^2} \left(\textrm{Im} G(\epsilon+sq) -\textrm{Im} G(\epsilon-sq) \right).
\end{eqnarray}
The r.h.s. of Eq.(\ref{deltaF}) is evaluated as: $\epsilon\ sQ_{max}\nu(1-\nu)\Delta(t) /E^2_{ph}$, where $Q_{max}\approx 2\mathcal{E}c(t)/s$ is the cutoff momentum corresponding to the crossover from the Lorentz solution to the self-consistent Born approximation solution. Then we solve Eq.(\ref{deltaF}) approximately as:
\begin{equation} \label{dfe}
\delta F(\epsilon)\sim -\frac{\nu(1-\nu)sQ_{max}}{E^2_{ph} \log(\left(\mathcal{E}c(t)/\Delta(t)\right)}\ \epsilon.
\end{equation}
This change of the electron distribution function is small $\delta F(sQ_{max})\ll 1$ because $sQ_{max}\ll \omega_H\ll E_{ph}$. Weakness of the electron phonon interaction means that during the period of \textit{ac}-field $1/\omega$ the variation of the electron distribution function due to the phonons is small: $dF(\epsilon)/dt\ll \omega$. Inserting Eq.(\ref{dfe}) into Eq.(\ref{Conductivity}) we still find zero in the leading logarithmic order but in the next order we find approximately
\begin{equation}
\sigma_{yy}\sim e^2N\nu(1-\nu)\omega_H/\tau E_{ph}^2.
\end{equation}
This conductivity is substantially smaller than the 'dark'-conductivity $\sigma_{yy}\sim e^2N$. Therefore driving electron system may result in that it becomes more ideal i.e. non-interacting with impurities system.

\section{Conclusion}

In conclusion we have found that a driven disordered 2DEG with the filling factor in high non-overlapping Landau levels remains in a nearly coherent quantum state with a broad non-Fermi-Dirac electron distribution function provided the electron-phonon interaction is sufficiently weak and the \textit{ac}-electric field is strong and slow. In this driven state the diagonal conductivity is strongly anisotropic with one component decreasing with increasing amplitude of the \textit{ac}-field and both being small as $\omega_H/\tau E^2_{ph}$. The \textit{ac}-field suppression of conductivity has a simple interpretation. Strong and slow \textit{ac}-electric field imparts a large velocity to an electron. Fast electron experiences a reduced scattering by disorder and behave like a free particle. In our theory the transport scattering time is equal to $\tau$ whereas in experiments it is 100 times larger. But qualitatively the experiment Ref.\onlinecite{Dorozhkin} in high mobility GaAs heterostructure has found a reduction of the $dc$-longitudinal conductivity in the magnetic fields above the cyclotron resonance $\omega_H>\omega$ that corresponds to slow \textit{ac}-fields and upon increasing the power of microwave irradiation.  

\begin{acknowledgments}
I would like to thank S.V. Iordanski for constant interest to this work and many useful suggestions. This work was supported by RFBR under the grant \#03-02-17229a.
\end{acknowledgments}


\begin{thebibliography}{22}
\bibitem{Mani} R. G. Mani, J. H. Smet \textit{et al}, Nature \textbf{420}, 646 (2002)
\bibitem{zdpw} M. A. Zudov, R. R. Du, L. N. Pfeiffer, and K. W. West, Phys. Rev. Lett. \textbf{90}, 046807 (2003); C. L. Yang, M. A. Zudov, T. A. Knuuttila, R. R. Du, L. N. Pfeiffer, and K. W. West, Phys. Rev. Lett. \textbf{91}, 096803 (2003)
\bibitem{ryzh} V. I. Ryzhii, JETP Lett., \textbf{7}, 28 (1968); V. I. Ryzhii, Sov. Phys. Solid State {\bf 9}, 2286 (1969)
\bibitem{dsrg} Adam C. Durst, Subir Sachdev, N. Read, and S. M. Girvin, Phys. Rev. Lett. \textbf{91}, 086803 (2003); J. Shi and X. C. Xie, Phys. Rev. Lett. \textbf{91}, 086801 (2003); I. A. Dmitriev, A. D. Mirlin and D. G. Polyakov, Phys. Rev. Lett. \textbf{91}, 226802 (2003)
\bibitem{ll03} X.L. Lei and S.Y. Liu, Phys. Rev. Lett., \textbf{91}, 226805 (2003)
\bibitem{ryzh03}  V. Ryzhii, Phys. Rev. B \textbf{68}, 193402 (2003)
\bibitem{AFS} T. Ando, A. B. Fowler, and F. Stern, Rev. Mod. Phys. \textbf{54}, 437-672 (1982)
\bibitem{PG} 'The Quantum Hall Effect', edited by E. R. Prange and S. M. Girvin ( Springer Verlag, New York, 1990)
\bibitem{Kohn} W. Kohn, Phys. Rev., {\bf 123}, 1242 (1961)
\bibitem{AGD} A. A. Abrikosov, L. P. Gor'kov and I. E. Dzyaloshinski, 'Methods of Quantum Field Theory in statistical Mechanics' ( Dover, New York 1963 )
\bibitem{Keldysh} L. V. Keldysh, Zh. Eksp. Teor. Fiz. \textbf{47}, 1515 (1964) [Sov. Phys. JETP \textbf{20}, 1018 (1965)]
\bibitem{gd91} F. Grossmann, T.Ditrich et al, Phys. Rev. Lett. \textbf{67}, 516 (1991)
\bibitem{Elias} G. M. Eliashberg, Zh. Eksp. Teor. Fiz. \textbf{41}, 1241 (1961) [Sov. Phys. JETP \textbf{14}, 886 (1962)]
\bibitem{bg61} Yu. A. Bychkov and L.P. Gor'kov, Zh. Eksp. Teor. Fiz. \textbf{41} 1592 (1961)
\bibitem{kubo65}  Kubo, R., S. J. Miyake, and N. Hashitsume, Solid State Phys. \textbf{17}, 269-364 (1965)
\bibitem{ryzhii} A. D. Malov and V. I.  Ryzhii, Sov. Phys. Solid State \textbf{14}, 1766 (1973)
\bibitem{Fermi} E. Fermi, Phys. Rev. \textbf{75}, 1169 (1949)
\bibitem{Dorozhkin} R. G. Mani, V. Narayanamurti, K. von Klitzing, J. H. Smet, W. B. Johnson and V. Umansky, 
Phys. Rev. B 69, 161306 (2004)
\end{thebibliography}
\end{document}